# Correcting for confounding in longitudinal experiments: positioning non-linear mixed effects modeling as implementation of standardization using latent conditional exchangeability.

Christian Bartels[1], Martina Scauda[2], Neva Coello[1], Thomas Dumortier[1], Bjoern Bornkamp[1], Giusi Moffa[3]

[1] Novartis Pharma AG, Basel, Switzerland

[2] Novartis Pharma AG, Basel, Switzerland; current affiliation: Statistical Laboratory, University of Cambridge, Cambridge, United Kingdom

[3] Dep. of Mathematics and Computer Science, University of Basel, Basel, Switzerland

**Correspondance :** Christian Bartels, Novartis Pharma AG, 4002 Basel, Switzerland, christian.bartels@novartis.com

## Abstract

Non-linear mixed effects modeling and simulation (NLME M&S) is evaluated to be used for standardization with longitudinal data in presence of confounders. Standardization is a well-known method in causal inference to correct for confounding by analyzing and combining results from subgroups of patients. The purpose of this paper is to show that non-linear mixed effects modeling is a particular implementation of standardization that conditions on individual parameters described by the random effects of the mixed effects model. The present work is motivated by the fact that in pharmacometrics NLME M&S is routinely used to analyze clinical trials and to predict and compare potential outcomes of the same patient population under different treatment regimens. Such a comparison is a causal question sometimes referred to as causal prediction. Nonetheless, NLME M&S is rarely positioned as a method for causal prediction.

As an example, a simulated clinical trial is used that assumes treatment confounder feedback in which early outcomes can cause deviations from the planned treatment schedule and are correlated with the final outcome due to their dependence on common individual parameters. Being interested in the outcome for the hypothetical situation that patients adhere to the planned treatment schedule, we put assumptions in a causal diagram. From the causal diagram, conditional independence assumptions are derived either using latent conditional exchangeability, conditioning on the individual parameters or using sequential conditional exchangeability, conditioning on earlier outcomes. Both conditional independencies can be used to estimate the estimand of interest, e.g., with standardization, and they give unbiased estimates. We will demonstrate this using a simulation-estimation experiment where assumptions required for identification of the estimand hold, including positivity, no unobserved confounding, and adequate specification of the models used for the analyses.

## Introduction

Mixed effects models are routinely used to analyze longitudinal clinical trial data[1, 2], however they are rarely discussed as method for causal inference in this setting[3, 4, 5]. Here we position mixed effects modeling as implementation of standardization (Section 13 and 21.1 of Ref[6]) using latent conditional exchangeability. We do this by looking at an example of treatment-confounder feedback, and derive and apply two sets of equations to evaluate the estimand of interest:

1. Standardization conditioning on early outcomes using sequential conditional exchangeability. This is one of the approaches proposed in textbooks on causal inference such as in part III section 21.1 of Ref[6].
2. Standardization conditioning on individual parameters, which one could refer to as using latent exchangeability. This corresponds to what is routinely done in pharmacometrics with population PK or PKPD modeling.

For the comparison of the two approaches, we state the estimand[7, 8] and formulate the assumptions in a causal diagram. From the causal diagram, conditional independence assumptions are derived. The conditional independencies can be used to estimate the estimand of interest, i.e., they identify the estimand. When using the methods to analyze data in a simulation-estimation experiment they both give valid estimates of the estimand and correct for the confounding. The methods and results sections give details. The remaining sections of the introduction provide some background and considerations on the example trial that we selected for illustration.

The motivation for the present work stems from the routine use of non-linear mixed effects modeling and simulation (NLME M&S) as an integral part of pharmacometrics [9, 10, 11] contributing to the analysis of clinical trials and to predict and compare potential outcomes of the same patient population under different hypothetical treatment regimens[12, 13]. Such a comparison is a causal question sometimes referred to as causal prediction[14]. The relation between pharmacometric analyses and causal inference has been discussed recently[13, 15, 16] without however considering the role of mixed effects modeling and the implied conditioning on latent variables.

Background

The pharmacokinetics and pharmacodynamics of many drugs and their effects on clinical endpoints can often be well described with pharmacometric non-linear mixed effects models[9, 10, 11]. These models consist of a non-linear structural model that describes the evolution of the pharmacokinetics and pharmacodynamics over time, and a mixed effects model that accounts for the differences between patient groups. Patients with similar values of the random effects have similar outcomes in that administering the drug will result in similar profiles of the response. The models are referred to as pharmacometric models since they are semi-mechanistic and based on pharmacological principles.

In practice, NLME M&S analyses are often executed by:

- First, fitting a non-linear mixed effects model to describe the observed data, i.e. the longitudinal response given the administered doses.
- Use model diagnostics to evaluate and ensure the quality of the description of the observed data.
- Finally, based on the fitted models, simulate response and evaluate alternative scenarios including different dosing regimens.

Formally, a generic mixed effect model can be represented as

$$E_{it} = f(x_{it}, \theta_i, b_i) + g(x_{it}, \theta_i, \xi)\varepsilon_{it}$$

$$\theta_i = h(\mu, \eta_i)$$

$$\eta_i \sim N(0, \Omega)$$

$$\varepsilon_{it} \sim N(0,1)$$

where:

- $E_{it}$ is the response at time $t$ in subject $i$.
- $f$ is referred to as the structural model and is a differentiable function of individual parameters, $\theta_i$, patient specific baseline covariates, $b_i$, and design variables $x_{it}$.
- $g$ is referred to as the residual error model with additional residual error parameters $\xi$, and with $\varepsilon_{it}$ being the independent and identically distributed residual error.
- $h$ is the model of the individual parameters that depends on population parameters, $\mu$, and patient specific random effects $\eta_i$.
- $\Omega$ is the covariance matrix of the distribution of the random effects.

For pharmacometric NLME models, the design variables usually contain the doses and the timing of the dose administrations and outcome assessments, e.g., $x_{it} = \{D_{it}, T_{it}, ...\}$ with $T_{it}$ being all the times of dose administrations and outcome observations of patient $i$ and $D_{it}$ all the administered drug amounts. Oftentimes the structural model, $f(x_{it}, \theta_i)$, is formulated as a set of first-order differential equations, the pharmacokinetics as a mammillary compartmental model and the pharmacodynamics as a direct or indirect response to the pharmacokinetics[17].

Standardization based on sequential conditional exchangeability is a method capable to adjust for time-dependent confounding even in the presence of treatment-confounder feedback (Section 21.1 of Ref[6]). Exchangeability refers to the concept that a subgroup of patients from a study who receive a particular treatment are representative of the overall study population. Randomization of treatment implies that treatment assignment and the potential outcomes under the different hypothetical treatments are exchangeable which helps answering many clinical questions of interest. When exchangeability holds, then the response observed in a subgroup gives an unbiased estimate of the response in the overall population. For some questions of interest, randomization is not sufficient to ensure exchangeability (see example discussed below) while conditional exchangeability may still hold. This means that exchangeability holds when restricting to patients with given characteristics which define a subset of the study population, e.g., all patients with the same age. It is then possible to estimate a treatment effect for each stratum. By averaging the effects over all strata, we can obtain the overall population effect. This is the idea of standardization, which extends to longitudinal trials thanks to a condition known as sequential conditional exchangeability.

In practice, analyses using standardization to perform causal inference are often executed by:

- Stating the question of interest by specifying treatment, population, outcome variable, population summary measure and intercurrent events, e.g., in the form of an estimand.
- Stating the assumptions on the data generating process, e.g., in the form of causal diagrams.
- Assess whether the estimand can be identified, i.e., whether it is possible, in principle and provided the assumptions are correct, to obtain a valid, unbiased estimate of the estimand of interest.
- Estimate the estimand of interest based on the available data using one of the methods for causal estimation such as standardization.

From the previous descriptions in this section, NLME M&S and standardization would appear as rather different approaches to analyze longitudinal data. In this work we aim to position NLME M&S as an implementation of standardization using latent conditional exchangeability; "latent" because the conditioning is on the non-directly observed individual PK parameters which are estimated from the observed data as part of the mixed effect model. As an implementation of standardization, NLME M&S is comparable to standardization using sequential conditional exchangeability. For certain situations, we find that they can be used almost interchangeably. For other situations, one or the other approach may be better suited, and we leave it to future work to demonstrate how the two approaches can complement each other.

### Considerations

For the derivations, we focus on one particular type of treatment-confounder feedback with assumptions presented in causal diagrams. To ensure a higher degree of generality we keep the diagrams and the derivations that they support generic without being specific on the nature of the interventions and the outcomes.

For illustration, we use a simulation with plasma concentrations of the drug over time as outcome (aka PK concentrations, PK exposure or simply exposure or PK in pharmacometrics). The aim is then to estimate the expected potential plasma concentrations under different treatment regimens. The rationale is that non-linear mixed effect models are routinely used in pharmacometrics to characterize PK, they tend to give precise descriptions of observed PK profiles and estimating individual PK parameters from the observed plasma concentration profiles is often quite reliable [9, 11, 18]. As such we can expect adjustments on individual PK parameters to work well and our simulation-estimation exercise is realistic. Further, characterizing PK after dose administration is a first step towards a more complete but also more complex characterization of the drug and its effects using non-linear mixed effects population PKPD models [10].

The chosen simulation scenario was inspired by a planned dose-finding study. The study evaluates different up-titration schemes aiming at identifying different treatment regimens with which most of the patients can rapidly reach the target dose. Due to the limited duration of the study, it is anticipated that patients may not be able to up-titrate because of initial high sensitivity to the drug after the first administrations. This high sensitivity is expected to disappear as patients continue to take the drug. In this setting estimating the outcome assuming that all patients can adhere to the up-titration schedule could be of interest, as it is considered probable that in longer phase 3 studies or in clinical practice all patients are ultimately able to up-titrate, but it may take longer for some of them (see also Refs[8, 19] for a discussion of the relevance of hypothetical estimands).

### Methods

For deriving the estimators, we consider the generic situation of a randomized longitudinal trial in which drug administration at later time points may be affected by the outcome at earlier time points. The drug gets administered repeatedly at different times $t$ and the outcome is measured before each drug administration. The randomization assigns a treatment strategy to each patient that defines the doses that should be taken at each time point. Intercurrent events may occur that affect the subsequent dosing so that it deviates from the randomly assigned treatment strategy. The occurrence of intercurrent events, such as insufficient efficacy or adverse events, is thought to be related to earlier outcomes. The interest lies in estimating the potential outcome under the hypothetical situation that no intercurrent event occurs and patients adhere to the randomly assigned treatment regimen. This corresponds to a hypothetical strategy as defined in the ICH E9R1 addendum[7]. Further, it is assumed that outcomes at different time points are correlated as they depend on the same set of individual parameters, and a mixed effects model adequately describes their statistical dependence. For expository purposes, we first present the derivation for a simpler case restricted to two dose administrations, and then extend to the general case of a longitudinal trial design with a larger number $n$ of dose administrations.

Directed acyclic graphs (DAGs) encode assumptions on experiments in the form of (conditional) independence relationships. The DAG in Figure 1 describes the assumptions for a trial with two administered doses. In the diagram, $R$ represents the randomization, $D_t$ the dose administrations at times $t$, $E_t$ the outcome just before the next dose at time $t+1$, $S_t$ the intercurrent event, and $\Theta$ the individual parameters. The outcome at time $t$ depends on all the doses at previous times $t' \leq t$. The same outcome has a causal effect on the intercurrent event at time $t$, while the doses depend on the randomization and prior intercurrent events. The individual parameters characterize the individual predisposition to certain outcomes at different times. As such they constitute common causes of all outcomes determining the association that exists between outcomes at different time points. Both, the individual parameters and the previous outcome are confounders for the effect of the intercurrent event on the subsequent outcome: the intercurrent event affects this outcome via a directed path mediated by the following dose, and a confounding pathway exists via the individual parameters and the previous outcome. The individual parameters and the previous outcome differ in that the outcome is directly observed whereas the individual parameters are estimated from the observed data in the mixed effects model of the NLME.

The estimand is the outcome $E_2$ in the study population for each treatment arm $r$ in a hypothetical scenario where the intercurrent event does not occur. Using the notation of counterfactuals, in a way similar to Ref[5], we want to intervene on the intercurrent event, so that $s_1 = 0$, and estimate the counterfactual distribution $Pr(E_2^{r,s_1=0})$. Single world intervention graphs (SWIGs) allow us to assess conditional exchangeabilities and establish whether the counterfactual distribution of interest is identifiable. Graphically conditional exchangeability holds if the target of intervention, $S_1^r$, and the outcome of interest, $E_2^{s_1=0}$, are d-separated given the set of covariates we can use for the adjustment (Section 7.5 of Ref[6]).

To obtain the SWIG corresponding to the DAG in Figure 1 we split the intervention nodes, $R$ and $S_1$, and use the counterfactual notation, e.g. $E_2^{r,s_1=0}$, for any outcomes downstream of the interventions. The independence of randomization and any potential outcomes, together with the consistency assumption, imply that $Pr(E_2^{r,s_1=0}) = \Pr(E_2^{s_1=0}|R=r)$. Because of the confounding paths through the individual parameters and the previous outcome, exchangeability does not hold between the potential outcome $E_2^{s_1=0}$ and the intercurrent event. Therefore, to estimate the counterfactual distribution from the observed data, we need to establish whether there are sets of observed variables with which conditional exchangeability between the potential outcome $E_2^{s_1=0}$ and the intercurrent event $S_1$ holds. Those so-called adjustment sets are such that they block all confounding paths when conditioned upon. An adjustment set from which no variable can be removed without losing conditional exchangeability constitutes a minimal adjustment set. The SWIG in Figure 1 has $\{E_1\}$ and $\{D_1, \Theta\}$ as the two minimally sufficient adjustment sets[20]. Both adjustment sets block all incoming paths that lead up to the intervention $S_1$. As such any other node can be added to the adjustment set without breaking the d-separation (see also Fine Point 6.1 of Ref[6]). Because adding conditioning variables on a path that is not uniquely blocked by a collider will not open it, the randomization $R$ can be added to the adjustment sets without breaking d-separation. Further, if we consider only treatment arms that assign the doses deterministically, then the randomization uniquely defines $D_1$ and we may remove $D_1$ from any adjustment set containing $R$. Therefore both $\{E_1, R\}$ and $\{R, \Theta\}$ constitute valid adjustment sets to derive equations for standardization: With the observed outcome $E_1$

$$\Pr(E_2^{S_1=0} | R = r) = \int_{e_1} Pr(E_2^{S_1=0} | R = r, E_1 = e_1) \Pr(E_1 = e_1 | R = r) de_1 \quad \text{by law of total probability}$$

$$= \int_{e_1} Pr(E_2^{S_1=0} | R = r, E_1 = e_1, S_1 = 0) \Pr(E_1 = e_1 | R = r) de_1 \quad \text{by conditional exchangeability}$$

$$= \int_{e_1} Pr(E_2 | R = r, E_1 = e_1, S_1 = 0) \Pr(E_1 = e_1 | R = r) de_1 \quad \text{by consistency.}$$

Similarly with the individual parameters $\Theta$

$$\Pr(E_2^{S_1=0} | R = r) = \int_\theta Pr(E_2^{S_1=0} | R = r, \Theta = \theta) \Pr(\Theta = \theta | R = r) d\theta \quad \text{by law of total probability}$$

$$= \int_\theta Pr(E_2^{S_1=0} | R = r, \Theta = \theta, S_1 = 0) \Pr(\Theta = \theta | R = r) d\theta \quad \text{by conditional exchangeability}$$

$$= \int_\theta Pr(E_2 | R = r, \Theta = \theta, S_1 = 0) \Pr(\Theta = \theta | R = r) d\theta \quad \text{by consistency}$$

$$= \int_\theta Pr(E_2 | R = r, \Theta = \theta, S_1 = 0) \Pr(\Theta = \theta) d\theta \quad \text{using independence of the individual parameters from the randomization.}$$

Both equations rely on the distribution of a confounder in the overall population estimated from the observed data and the observed conditional distribution of the outcome given the intervention $S_1 = s_1$ and the confounder. The confounder is either the previous outcome $\Pr(E_1 = e_1 | R = r)$ conditional on the treatment arm $r$ or the individual parameters $\Pr(\Theta = \theta)$. The conditional distributions are $\Pr(E_2 | R = r, \Theta = \theta, S_1 = s_1)$ or $Pr(E_2 | R = r, E_1 = e_1, S_1 = s_1)$, respectively. To describe these distributions, we use parametric models which we estimate from the observed data. The fitted models characterize the distribution of the counterfactual outcomes $E_2^{S_1}$ in that they can be used to sample counterfactuals under different treatment strategies (Section 21.1 of Ref[6]).

Conditioning on the previous outcome $E_1$ corresponds to standardization as described in Ref[5]. Conditioning on the individual PK parameters differs in that the parameters are not directly observed but estimated by the random effects of a (non-linear) mixed effects model. The mixed effects model consists of the model of the distribution of the individual parameters $\Pr(\Theta = \theta)$ and the conditional model of the outcome $Pr(E_2 | R = r, \Theta = \theta, S_1 = s_1)$ given the individual parameters and the treatments. In the NLME notation of the background section, they correspond to $h(.)$, $\Omega$, and $f(.)$, respectively. With the NLME model fitted to the observed data, we can then simulate the counterfactual outcomes[16].

*Figure 1. DAG and SWIG for 2 doses.*

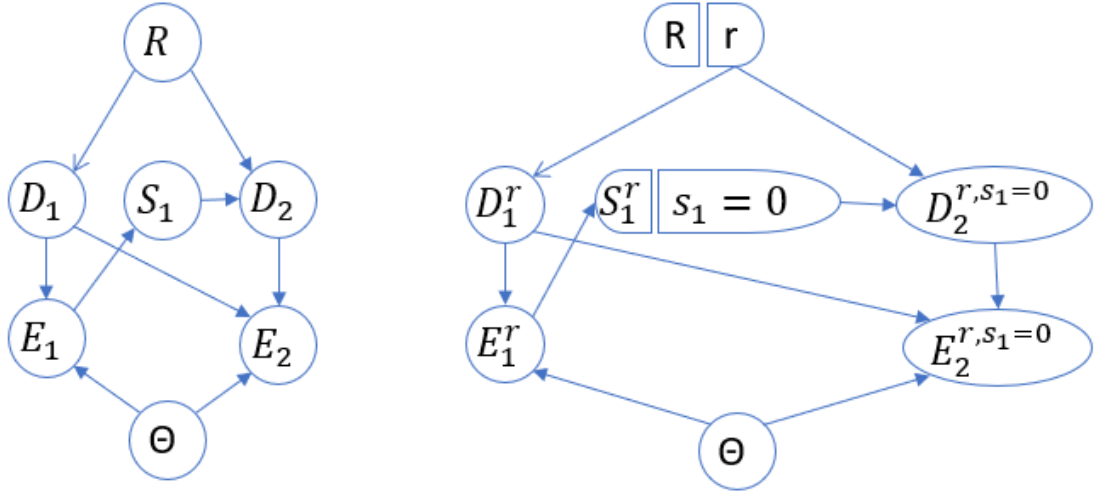

R, randomization; $D_t$ the dose administration at times $t$; $E_t$, outcome at the end of the dosing interval starting at time $t$; $S_t$, intercurrent event; Θ, individual parameters. In the SWIG, lower case is used to denote the intervention of interest.

Going beyond two doses, the SWIG in Figure 2 represents a more general longitudinal trial with several administrations, where we assume interventions on the randomization and all the intercurrent events. $E_t^{r,\overline{s_{t-1}}=0}$ refers to the counterfactual outcome at time $t$ if intervening on the randomization $r$ and on all intercurrent events $\overline{s_{t-1}}$ prior to $t$. Quantities with an overbar and a time subscript, e.g., $\overline{E_t^{r,s=0}}$, refer to all values up to and including time t; quantities with an overbar but without a time subscript, e.g., $\overline{E^{r,s=0}}$, refer to the values for all time points. For all counterfactuals it is assumed that they depend on all prior interventions through the mediation of prior administered doses without making this explicit in the notation. An underbar, e.g., $\underline{E_\tau^{r,\overline{s_t}}}$, is used to indicate all values at or after the indicated time. Our objective is to estimate the counterfactual distribution $Pr\big(E_n^{r,\overline{s_{n-1}}=0}\big)$ at the last time point $n$, which is equal to $Pr\big(E_n^{r,\overline{s_{n-1}}=0}|R=r\big)$, thanks to the independence of the potential outcomes from randomization.

To derive estimation equations we follow arguments as in Technical Point 19.3 of Ref[6]. In the longitudinal setting of the SWIG in Figure 2, sequential conditional exchangeability holds for any given intercurrent event with observations (outcomes and doses) that occurr after the event conditional on all prior observations and intercurrent events, i.e., for $\tau > t$

$$\left(\underline{D_\tau^{r,\overline{s_t}=0}}, \underline{E_\tau^{r,\overline{s_t}=0}}\right) \perp S_t^{r,\overline{s_{t-1}}=0} | \overline{E_t^{r,s=0}}, \overline{D_t^{r,s=0}}, \overline{S_{t-1}^{r,s=0}}, \theta, R=r$$

The conditioning is redundant, in that similarly to the case with two doses, either the outcomes $\overline{E_t^{r,s}}$ or the doses and the random effects, $\overline{D_t^{r,s}}$ and $\theta$, can be removed from the adjustment set without opening confounding paths into $S_t^{r,\overline{s_{t-1}}}$. Focusing on the set without prior outcomes, we have

$$\left(\underline{D_\tau^{r,\overline{s_t}=0}}, \underline{E_\tau^{r,\overline{s_t}=0}}\right) \perp S_t^{r,\overline{s_{t-1}}=0} | \overline{D_t^{r,s=0}}, \overline{S_{t-1}^{r,s=0}}, \theta, R=r$$

Since in the example the doses are determined by the randomization and intercurrent event, it is redundant to condition on all three of them. The doses can be removed, which gives

$$\left(D_\tau^{r,\bar{s}_t=0}, E_\tau^{r,\bar{s}_t=0}\right) \perp S_t^{r,\overline{s_{t-1}}=0} \mid \overline{S_{t-1}^{r,s=0}}, \theta, R = r$$

Using this exchangeability repeatedly from time 1 to $t$ enables estimation of the countefactual distribution from the observed data. The $S_t^{r,\overline{s_{t-1}}=0}$ for $t = 1 \dots n-1$ enter the conditioning set sequentially, are set to the target intevention 0, and are combined into $\overline{S_{n-1}^{r,s=0}} = 0$:

$Pr\left(E_n^{r,\overline{s_{n-1}}=0} \mid R = r\right) = \int_\theta Pr\left(E_n^{r,\overline{s_{n-1}}=0} \mid R = r, \Theta = \theta\right) \Pr(\Theta = \theta \mid R = r) \, d\theta$

by law of total probability

$= \int_\theta Pr\left(E_n^{r,\overline{s_{n-1}}=0} \mid R = r, \Theta = \theta\right) \Pr(\Theta = \theta) \, d\theta$

since the random effects are independent of randomization

$= \int_\theta Pr\left(E_n^{r,\overline{s_{n-1}}=0} \mid R = r, \overline{S_1^{r,s=0}} = 0, \Theta = \theta\right) \Pr(\Theta = \theta) \, d\theta$

by conditional exchangeability for time point 1

$= \int_\theta Pr\left(E_n^{r,\overline{s_{n-1}}=0} \mid R = r, \overline{S_{n-1}^{r,s=0}} = 0, \Theta = \theta\right) \Pr(\Theta = \theta) \, d\theta$

by repeated application of conditional exchangeability from time 2 to $n-1$

$= \int_\theta Pr(E_n \mid R = r, \overline{S_{n-1}} = 0, \Theta = \theta) \Pr(\Theta = \theta) \, d\theta$

by consistency

*Figure 2. SWIG for longitudinal trial.*

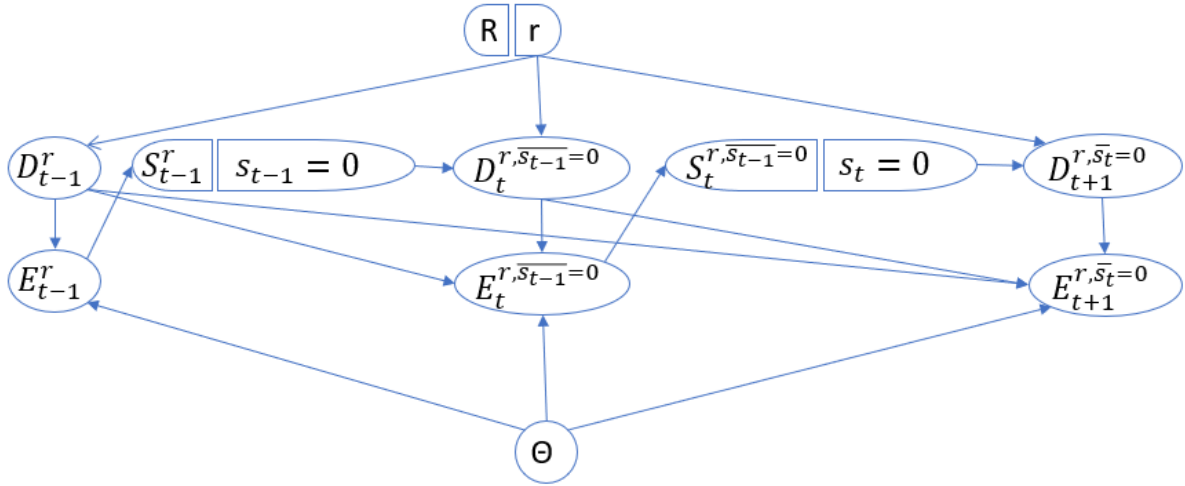

R, randomization; $D_t$ the dose administration at times $t$; $E_t$, outcome at the end of the dosing interval starting at time $t$; $S_t$, intercurrent event; $\Theta$, individual parameters. Lower case is used to denote the intervention of interest.

The derivation conditioning on earlier outcomes is similar, except that at each time point, earlier exposures need to be added to the conditioning set using the law of total probability e.g., see also Chapter

III and section 19.3 of Ref[6] or Ref[5]. When removing the redundant doses and random effects from the conditioning set, while retainining the earlier outcome, sequential exchangeability takes the form

$$\left(D_\tau^{r,\bar{S}_t=0}, E_\tau^{r,\bar{S}_t=0}\right) \perp S_t^{r,\bar{S}_{t-1}=0} | \overline{E_t^{r,S=0}}, \overline{S_{t-1}^{r,S=0}} = 0, R = r$$

Based on consistency (Technical Point 19.3 of Ref[6]), counterfactuals on the right side can be replaced giving

$$\left(D_\tau^{r,\bar{S}_t=0}, E_\tau^{r,\bar{S}_t=0}\right) \perp S_t | \overline{E_t}, \overline{S_{t-1}} = 0, R = r$$

Starting from this, we have for the first intercurrent event

$$Pr\left(E_n^{r,\overline{S_{n-1}}=0} | R = r\right) = \int_{e1} Pr\left(E_n^{r,\overline{S_{n-1}}=0} | R = r, E_1 = e_1\right) Pr\left(E_1 = e_1 | R = r\right) de_1$$

by law of total probability

$$= \int_{e1} Pr\left(E_{t+1}^{r,\overline{S_{n-1}}=0} | R = r, S_1 = 0, E_1 = e_1\right) Pr\left(E_1 = e_1 | R = r\right) de_1$$

by conditional exchangeability for time $t = 1$.

Then for the second intercurrent event

$$Pr\left(E_n^{r,\overline{S_{n-1}}=0} | R = r\right) =$$ 

by law of total probability

$$\int_{e1,e2} Pr\left(E_n^{r,\overline{S_{n-1}}=0} | R = r, S_1 = 0, E_1 = e_1, E_2 = e_2\right) \times$$
$$Pr\left(E_1 = e_1 | R = r\right) \times$$
$$Pr\left(E_2 = e_2 | R = r, S_1 = 0, E_1 = e_1\right) de_1 de_2$$

$$Pr\left(E_n^{r,\overline{S_{n-1}}=0} | R = r\right) =$$ 

by conditional exchangeability for

$$\int_{e1,e2} Pr\left(E_n^{r,\overline{S_{n-1}}=0} | R = r, S_1 = 0, S_2 = 0, E_1 =, E_2 =\right) \times \quad \text{time } t = 2$$
$$Pr\left(E_1 = e_1 | R = r\right) \times$$
$$Pr\left(E_2 = e_2 | R = r, S_1 = 0, E_1 = e_1\right) de_1 de_2$$

Repeating this for all intercurrent event up to time $n - 1$ and using the convention that terms with time index 0, i.e., $E_0^{S=0}$, can be removed, gives

$$Pr\left(E_n^{r,\overline{S_{n-1}}=0} | R = r\right) =$$
$$\int_{\overline{e_{n-1}}} Pr\left(E_n^{r,\overline{S_{n-1}}=0} | R = r, \overline{S_{n-1}} = 0, \overline{E_{n-1}} = \overline{e_{n-1}}\right) \times$$
$$\prod_{t=1}^{n-1} Pr\left(E_t | R = r, \overline{S_{t-1}} = 0, \overline{E_{t-1}} = \overline{e_{t-1}}\right) d\overline{e_{n-1}}$$

And finally by consistency

$$Pr\left(E_n^{r,\overline{S_{n-1}}=0} | R = r\right) = \int_{\overline{e_{n-1}}} \prod_{t=1}^{n} Pr\left(E_t | R = r, \overline{S_{t-1}} = 0, \overline{E_{t-1}} = \overline{e_{t-1}}\right) d\overline{e_{n-1}}$$

Both variants of the estimation equations, the one conditioning on the random effects or the one conditioning on earlier outcomes, can again be implemented using a parametric modeling and simulation approach. Estimation conditioning on earlier outcomes requires models $Pr(E_t | R = r, \overline{E_{t-1}}, \overline{S_{t-1}} = 0)$ for all times $t$. Conditioning on the random effects requires the conditional model for the outcome, $Pr(E_n|R = r, \bar{S} = 0, \Theta = \theta)$, and the model of the distribution of the random effect in the overall population, $Pr(\Theta = \theta)$. With the deterministic relation of the doses given randomization and intercurrent event, the model of the outcome can be formulated as conditioning on the doses, $Pr(E_n|\bar{D} = \bar{d}, \Theta = \theta)$, which is the variant that is commonly used in pharmacometrics NLME M&S.

In both cases, the models are estimated from the observed data. The fitted models are used to simulate a large number, e.g., 5000, of counterfactual outcomes $E_n^{r,\overline{S_{n-1}}=0}$ that represent the distribution of interest.

## Results

### Simulation of a trial with treatment confounder feedback

To evaluate the different estimation methods, a trial was simulated assuming a treatment-confounder feedback as illustrated in the causal graph (Figure 2) with PK exposure being the outcome, previous exposure having a direct effect on the intercurrent event and the exposures at different time points being correlated due to their dependence on common individual parameters. This simulated trial constitutes the observed data to be analyzed with the different standardization approaches. The study evaluates five treatment arms with different target doses and different up-titration schedules (Table 1). As outcome of interest, we focused on exposure at week 8, two weeks after the last dose.

*Table 1: Treatment arms of the simulated study.*

| ARM | DESCRIPTION | D1 | W2 | W4 | W6 |
|---|---|---|---|---|---|
| 1 | Target dose 60 | 30 | 60 | 60 | 60 |
| 2 | Target dose 120, up-titration (a) | 30 | 60 | 120 | 120 |
| 3 | Target dose 120, up-titration (b) | 60 | 120 | 120 | 120 |
| 4 | Target dose 240, up-titration (a) | 60 | 120 | 240 | 240 |
| 5 | Target dose 240, up-titration (b) | 60 | 240 | 240 | 240 |

*Doses (mg) given at day 1, and then after 2, 4 or 6 weeks.*

Each treatment arm was simulated to consist of 5000 patients. The large number of patients was selected to reduce the statistical uncertainty in the estimates and facilitate the detection of causal biases in the estimation methods. A 1-compartment PK model with bolus intravenous administration which assumes a linear elimination was used for the simulations [21, 22]. The model includes random effects to estimate individual's clearances and volumes. The random effects were considered to be independent of each other. A proportional residual error model was used.

In terms of the generic NLME from the introduction, the design variables of the example consist of the doses and the timing of the doses and PK sampling, $x_{it} = \{D_{it}, T_{it}\}$; to simplify we assume that taking the trough sample and administering the next dose happen immediately after each other at the same time $T_{it}$. The one compartmental PK model corresponds to the structural model and can be written as a sum of exponentials, e.g., for trough exposure after $n$ dose administrations as

$$f(x_{it}, \psi_i) = \frac{1}{v_i}\sum_{t=1}^{n} D_{it} e^{-cl_i/v_i (T_{i(n+1)} - T_{it})},$$

where the individual parameters $\theta_i$ consist of the individual apparent clearance, $cl_i$, and apparent volume, $v_i$. The simulation model assumed independent log-normal distributions for the individual parameters, e.g., $cl_i = \exp(\mu_{cl} + \eta_{cl,i})$ with $\eta_{cl,i} \sim N(0, \omega_{cl})$ and $\mu_{cl}$ the fixed effect model parameter and $\omega_{cl}$ the standard deviation of the random effect. The assumed proportional residual error is $g(x_{it}, \psi_i, \xi)\varepsilon_{it} = f(x_{it}, \psi_i)\,\xi\,\varepsilon_{it}$.

The parameters are listed in Table 2, and the resulting PK profiles are shown in Figure 3.

*Table 2: Parameters used for simulations.*

| Parameter | Value |
|---|---|
| **Structural parameters** | |
| Clearance (L/h), $\mu_{cl}$ | 0.0025 |
| Volume (L), $\mu_v$ | 2 |
| **Between-subject variability, SD (random effect)** | |
| BSV on clearance, $\omega_{cl}$ | 0.3 |
| BSV on volume, $\omega_v$ | 0.3 |
| **Residual variability, SD** | |
| Proportional, $\xi$ | 0.02 |

*PK model parameters used for simulations; symbols and formula are explained in the text.*

*Figure 3. Plasma concentration time profiles of the different treatment arms of the simulated study.*

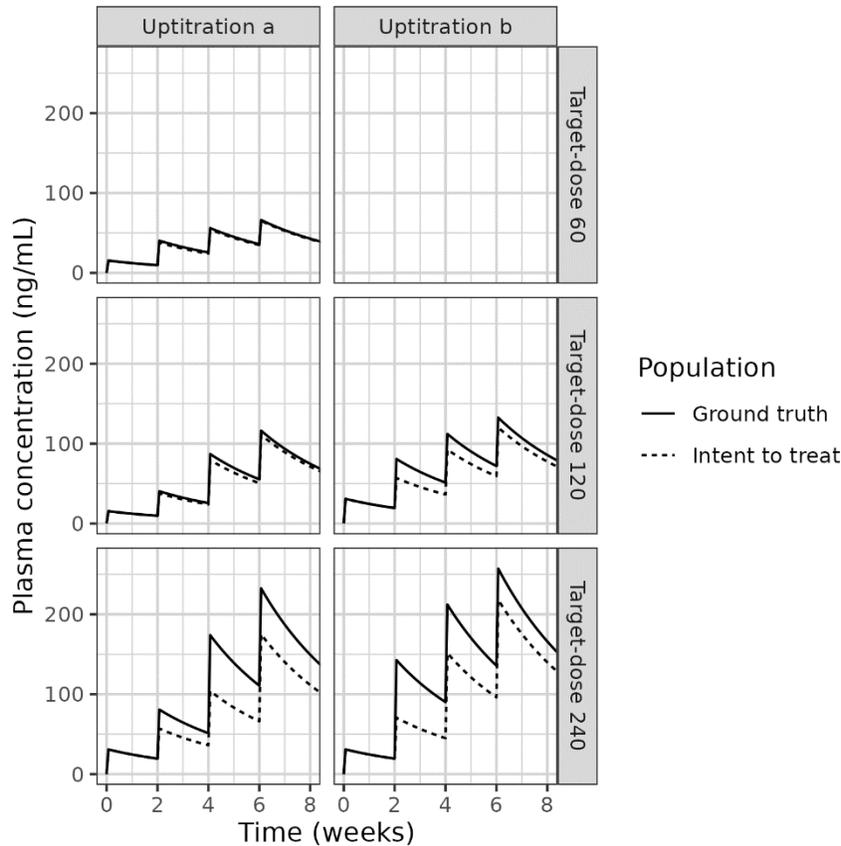

*Simulated profiles of plasma mean concentrations before (Ground truth) and after (Intent to treat) introducing the treatment-confounder feedback in the simulations. Target-dose and up-titration as defined in Table 1.*

Treatment-confounder feedback was implemented with high exposure being associated with increased probability of adverse events leading to delayed up-titration (e.g., patients with adverse events stay on the same dose level rather than up-titrating as planned). The probability of nonadherence, $p_S$, was assumed to be

$$p_S = \text{inv.logit}\left(\beta \log\left(\frac{E_t}{\alpha_t}\right)\right).$$

The thresholds were set to increasing values $\bar{\alpha} = (\alpha_1, \alpha_2, \alpha_3) = (15, 40, 100)$ ng/mL at weeks 2, 4 and 6, respectively, and chosen to correspond to typical exposures at these times. The parameter $\beta$, which determines the strength of the association, was set to 5. These parameters were selected so that the treatment-confounder feedback had a clear impact on the outcome, e.g., an about 14% reduction in exposure for treatment arm 5 (target dose 240, up-titration b). Figure 4 depicts the resulting trough exposures at week 8 after the fourth dose for treatment arm 5 (target dose 240, up-titration b). The plot shows the distribution of exposures that would be observed with full adherence (Ground truth), and the distributions when there is treatment-confounder feedback in the overall population (Intent to treat), and in the subset of patients that were able to adhere to the protocol (Per protocol), respectively. The exposure in the intent to treat population is smaller than the ground truth, since some patients do not up-titrate and receive lower doses. The exposure in the per protocol population is lower than the ground truth, since it summarizes patients that have a propensity for low exposure, since these are the patients that are most likely to adhere.

*Figure 4. Different summaries of plasma concentrations for one treatment arm.*

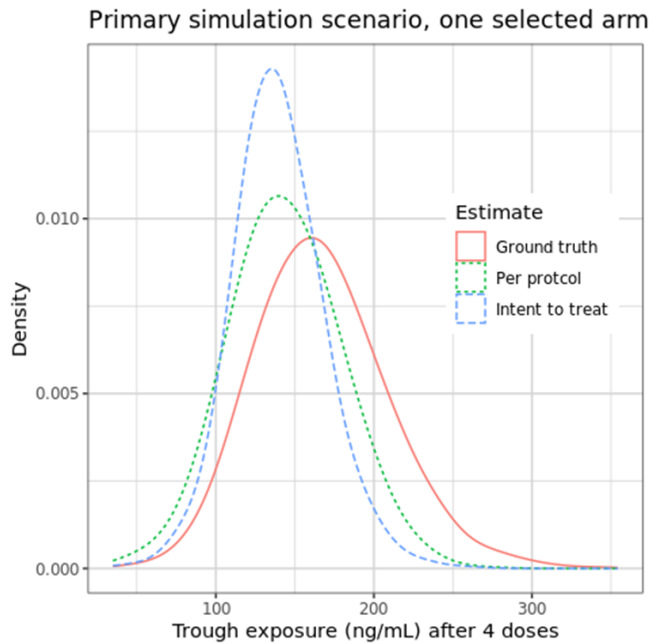

*Distribution of trough exposures at week 8 after the fourth dose for treatment arm 5 (target dose 240, up-titration b). Ground truth, result known from simulation setup without treatment-confounder feedback; per protocol, summary of patients that adhere to assigned treatment; intent to treat, summary of all patients.*

Implementation of the estimation methods

The following subsections describe the implementation of the estimation methods. Focus is on standardization conditioning either on early exposure or individual PK parameters as derived in the methods section. Inverse probability weighting (IPW)[6] is proposed and evaluated as alternative. Collectively these estimation methods are referred to as g-methods.

*Standardization conditioning on early exposure*

Parametric standardization conditioning on previous exposures requires models for the conditional distribution of the outcome given the previous exposures $Pr(\overline{E_t}|R = r, \overline{E_{t-1}}, \overline{S_{t-1}} = 0)$. Note that the interventions, i.e., randomization and the intercurrent events, define the doses, and we use the doses rather than the randomization or intercurrent events to specify the models (see also Discussion).

For the first exposure we assume that the logarithm of the dose normalized exposure is normally distributed with location parameter $\beta_0$ and scale parameter $\gamma_0$, i.e., as

$$log\left(\frac{E_1}{D_1}\right) = \beta_0 + \gamma_0 \varepsilon.$$

Here and subsequently $\varepsilon \sim N(0,1)$ is a standard normal distribution. For the conditional model of subsequent exposures, we also assume a normal distribution. Its location depends on prior doses and the dose normalized exposure after the first dose $E_{norm} = E_1/D_1$, and its scale on the previous two doses, i.e., as

$$E_t = \beta_{t,0} + \sum_{\tau=1}^{t} \beta_{t,\tau} D_\tau E_{norm} + \gamma_{D_{t-1}, D_t}$$

with the $\beta$ and $\gamma$ being model parameters and more specifically $\gamma_{D_{t-1}, D_t}$ being different scale parameters for each combination of prior doses. Both models were evaluated using model diagnostics that indicated an appropriate fit of the observed data.

*Using NLME to condition on individual PK parameters*

Parametric standardization conditioning on individual PK parameters, $\Theta$, requires models for the distribution of the individual PK parameters $Pr(\Theta = \theta)$ and a conditional model for the outcome given individual PK parameters $Pr(\overline{E}|\overline{D}, \Theta = \theta)$. As introduced in the background section, both are components of (non-linear) mixed effects models, i.e., $h(.)$, $f(.)$ and $\Omega$, and can be estimated together by fitting the models to the observed data. We used the same parametric model as was used for the simulation of the trial (Table 2). Note again that the randomization and the intercurrent events together define the doses that a patient receives, and that as is commonly done in pharmacometrics NLME M&S doses were used for the model fitting.

*Using inverse probability weighting (IPW) conditioning on previous exposures*

As an alternative to standardization, we also implemented inverse probability weighting (IPW) as described in text books e.g., section 21.2 of Ref[6]. IPW requires a model of the probability of the intercurrent events given previous exposures. We use the same relation as was used to introduce the treatment-confounder feedback in the simulated trial.

Estimation using standardization

This section presents the estimates obtained for the hypothetical scenario that the intercurrent event does not occur for the simulated trial with treatment-confounder feedback. This scenario is known to us from the simulation setup, and we refer to it as ground truth (Figure 4). Being interested in the exposure for the planned treatment schedule, we evaluate possibilities to correct for the confounding using NLME M&S or other g-methods (inverse probability weighting (IPW), standardization) for causal inference. We find that all methods can correct for the confounding, as long as the identifiability assumptions required for valid causal inference hold, including in particular positivity, no unobserved confounding, and correct specification of parametric models.

Figures 5 and 6 summarize the results. Figure 5 shows trough exposures after the fourth dose for treatment arm 5 (target dose 240, up-titration b). The plot displays the distribution of exposures that would be observed under full adherence (Ground truth) as reference, and the distributions in the presence of treatment-confounder feedback in the overall population of the simulated trial (Intent to treat), and in the subset of patients that could adhere to the protocol (Per protocol), respectively. The ground truth represents our main estimand, which as outlined in the introduction, is of interest, since it represents the situation that all patients reach the target dose. Reaching the target dose is an important outcome because it reflects what could happen in longer trials or in clinical practice when patients have more time to up-titrate. The intent to treat aims at a different estimand. It is of interest here since its difference to the ground truth illustrates the magnitude of the confounding due to the treatment-confounder feedback and since it can also be seen as an estimate of our estimand of interest that fails to adjust for the confounding. With the per protocol analysis, we simply illustrate a suboptimal estimate of our main estimand of interest; it corresponds to the result of inverse probability weighting with all weights being equal. Thus, these three distributions provide a reference of what we aim at estimating (ground truth) and two erroneous estimates of it that fail to adjust for the confounding (per protocol and intent to treat). Relative to the suboptimal estimates represented by the per protocol and intend to treat analyses, all estimation methods approach the ground truth both as judged by the location and the scale of the distribution.

Figure 6 summarizes results across treatment arms for all estimation methods and lists the mean and standard deviations of the distributions. To facilitate comparison with the ground truth, it also lists the mean and standard deviation relative to those of the ground truth. When looking at these numbers, it appears that NLME performs better than the other estimation methods. This is an artifact, however, from the setup of our simulation-estimation experiment. For NLME, we have used the same parametric model for the analysis as that used to simulate the trial data. This contrasts with standardization which relied on generic regression models to fit the simulated data. This approximation contributes to the (small) differences between the estimates obtained with these methods relative to the ground truth. For the IPW method, we calculated the weights using the same model as that used in the simulations for the probability of the intercurrent event. However, IPW also suffers from lack of positivity, since we also chose to parametrize the treatment-confounder feedback in a way to introduce an important difference between the distribution observed in the per-protocol population and that of interest (ground truth). This means that estimating the distribution for high exposures of interest will be challenging, because there will remain very few patients in the per protocol population.

*Figure 5. NLME M&S and other g-methods do provide a good estimate of the ground truth.*

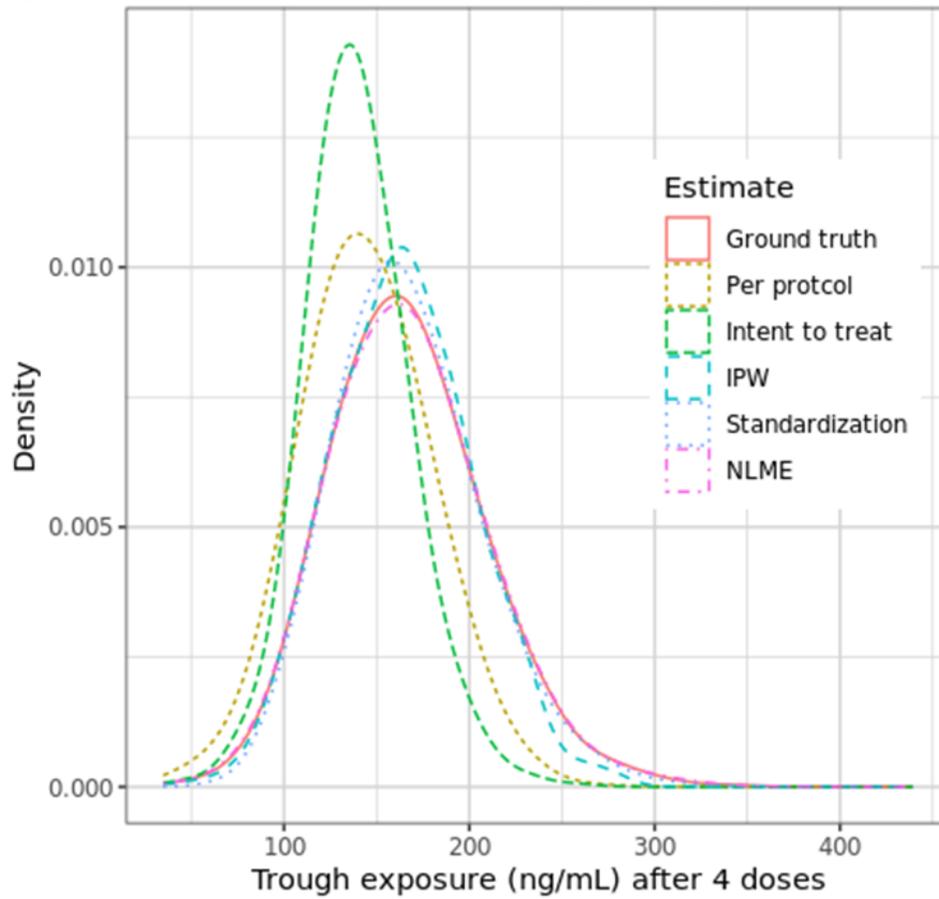

*Distribution of trough exposures at week 8 after the fourth dose for treatment arm 5 (target dose 240, up-titration b). Ground truth, per protocol, and intent to treat, as reference. Standardization, standardization conditioning on previous exposures; NLME, standardization conditioning on individual parameters; IPW, inverse probability weighting.*

*Figure 6. Comparison of estimation methods for the different treatment arms.*

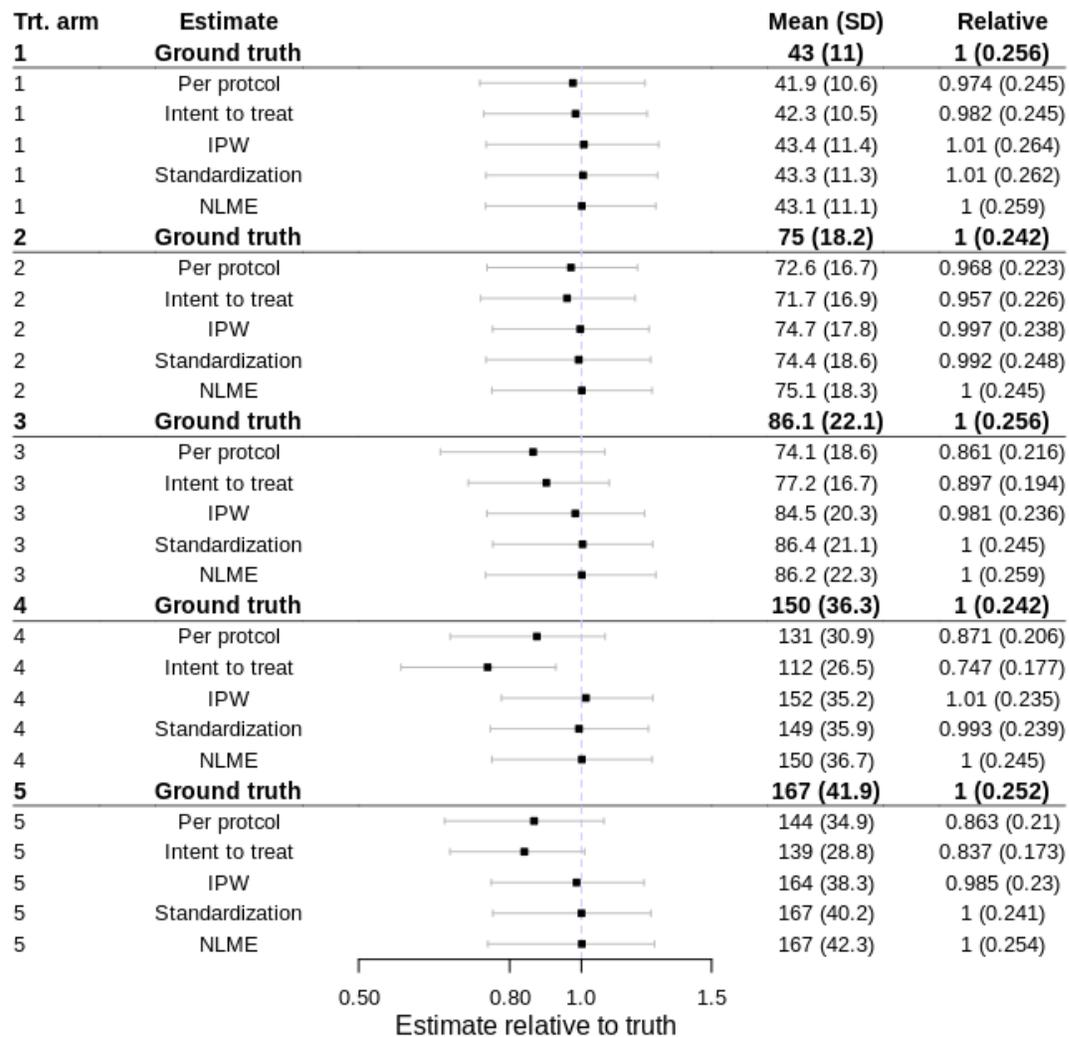

| Trt. arm | Estimate | Mean (SD) | Relative |
|---|---|---|---|
| **1** | **Ground truth** | **43 (11)** | **1 (0.256)** |
| 1 | Per protocol | 41.9 (10.6) | 0.974 (0.245) |
| 1 | Intent to treat | 42.3 (10.5) | 0.982 (0.245) |
| 1 | IPW | 43.4 (11.4) | 1.01 (0.264) |
| 1 | Standardization | 43.3 (11.3) | 1.01 (0.262) |
| 1 | NLME | 43.1 (11.1) | 1 (0.259) |
| **2** | **Ground truth** | **75 (18.2)** | **1 (0.242)** |
| 2 | Per protocol | 72.6 (16.7) | 0.968 (0.223) |
| 2 | Intent to treat | 71.7 (16.9) | 0.957 (0.226) |
| 2 | IPW | 74.7 (17.8) | 0.997 (0.238) |
| 2 | Standardization | 74.4 (18.6) | 0.992 (0.248) |
| 2 | NLME | 75.1 (18.3) | 1 (0.245) |
| **3** | **Ground truth** | **86.1 (22.1)** | **1 (0.256)** |
| 3 | Per protocol | 74.1 (18.6) | 0.861 (0.216) |
| 3 | Intent to treat | 77.2 (16.7) | 0.897 (0.194) |
| 3 | IPW | 84.5 (20.3) | 0.981 (0.236) |
| 3 | Standardization | 86.4 (21.1) | 1 (0.245) |
| 3 | NLME | 86.2 (22.3) | 1 (0.259) |
| **4** | **Ground truth** | **150 (36.3)** | **1 (0.242)** |
| 4 | Per protocol | 131 (30.9) | 0.871 (0.206) |
| 4 | Intent to treat | 112 (26.5) | 0.747 (0.177) |
| 4 | IPW | 152 (35.2) | 1.01 (0.235) |
| 4 | Standardization | 149 (35.9) | 0.993 (0.239) |
| 4 | NLME | 150 (36.7) | 1 (0.245) |
| **5** | **Ground truth** | **167 (41.9)** | **1 (0.252)** |
| 5 | Per protocol | 144 (34.9) | 0.863 (0.21) |
| 5 | Intent to treat | 139 (28.8) | 0.837 (0.173) |
| 5 | IPW | 164 (38.3) | 0.985 (0.23) |
| 5 | Standardization | 167 (40.2) | 1 (0.241) |
| 5 | NLME | 167 (42.3) | 1 (0.254) |

*Mean and standard deviation (SD) of trough exposures at week 8 after the fourth dose for the different treatment arms (Table 1). Ground truth and estimates: Standardization, standardization conditioning on previous exposures; NLME, standardization conditioning on individual parameters; IPW, inverse probability weighting. The relative results are the mean and SD of the estimated values divided by the mean of the ground truth.*

## Discussion

Causal inference methods rely on identifiability assumptions, in particular, exchangeability (or no unobserved confounding), consistency, positivity and correct specification of parametric models. For the present illustration, we have purposely chosen a ground truth scenario largely fulfilling the necessary assumptions. There was no unobserved confounding, positivity was maintained in an adequate range, and parametric models were either exactly matching the ground truth or provided a good approximation. In the presented evaluation using simulation-estimation with known ground truth, any of the methods can be made to fail.

From the point of view of evaluating the population PK approach, we noted that results were questionable when continuing to use a simple linear PK model for the analysis but change the ground truth and simulate the study data with non-linear PK assuming target mediated deposition (Figure 7, scenario labelled nonLinear); see, e.g., Ref[22] for an overview over some PK models. This would be a typical case of model misspecification (using the wrong parametric model), which in practice we can detect with the help of model diagnostics. Population PK results didn't change much when increasing the degree of confounding (more IEs) or the magnitude of the residual error (Figure 7, scenario labelled highRes). Interestingly, all estimation methods performed reasonably well against attempts to introduce unobserved confounding. In this attempt, we simulated IEs with a probability dependent on the unobserved maximal PK concentration rather than observed trough concentrations. All the estimation methods continued to provide reasonable estimates of the ground truth (Figure 7, scenario labelled cmax). In our setup the trough exposure depends on the individual PK parameters, both on the individual volume and the individual clearance, whereas the maximal exposure depends only on the individual volume. Thus, conditioning on the individual volume and clearance using NLME continues to block all confounding paths in this situation, and the NLME is expected to remain valid. Conditioning on prior trough exposure does not block all paths, and some bias in the estimates is expected. Nonetheless trough exposures are a proxy for maximal exposure up to the addition of random variability due to the variability of the clearance, and conditioning on prior trough exposure does help to correct for some of the confounding.

When assessing NLME as an implementation of causal standardization, it seems advantageous to use an example for which NLME models are well established. Here we looked at drug pharmacokinetic exposure (drug plasma concentrations) as outcome. The existence of well-established population PK models (NLME) for the PK exposure[21, 22] motivated our choice. Available PK models often describe the data well and the available data is often sufficient to enable reliable estimation of individual parameters.

Future work should also look at treatment-confounder feedback with efficacy or safety outcomes as the focus. One may want to differentiate three situations. First, if there is sufficiently rich longitudinal efficacy data, all the approaches presented here including NLME could work. Second, if the intercurrent event does not depend on efficacy, e.g., it is driven by drug exposure and tolerability, then efficacy is not a confounder. The approaches proposed and evaluated here can be expanded by adding an exposure-response component such as going to PKPD models. Third, if we have only limited longitudinal efficacy data such as in a trial evaluating time to a terminal event, and the IE nonetheless depends on some unobserved aspects of efficacy, then none of the discussed approaches seems to provide a solution.

We found that it was difficult to determine adequate parametric models for the different estimation methods, except when we knew the correct model from the simulations that generated the trial data. This was the case for the probability of nonadherence, which we used for the IPW approach, and in the models for standardization when conditioning on individual PK parameters. For the models for standardization conditioning on early exposure, we ended up using linear models with the dose normalized exposure after the first dose as regressor for the location of the conditional distribution. This choice was based on the knowledge that we have linear dose proportional PK, which we knew from the simulation setup. Also, model diagnostics can support the selection of an appropriate parametric model, and their use is common practice in pharmacometrics analyses of clinical data with NLMEs. Alternatively, machine learning or other non-parametric methods could be envisaged and may avoid the need of having to identify appropriate parametric models.

*Figure 7. Comparison of different g-methods for different types of data.*

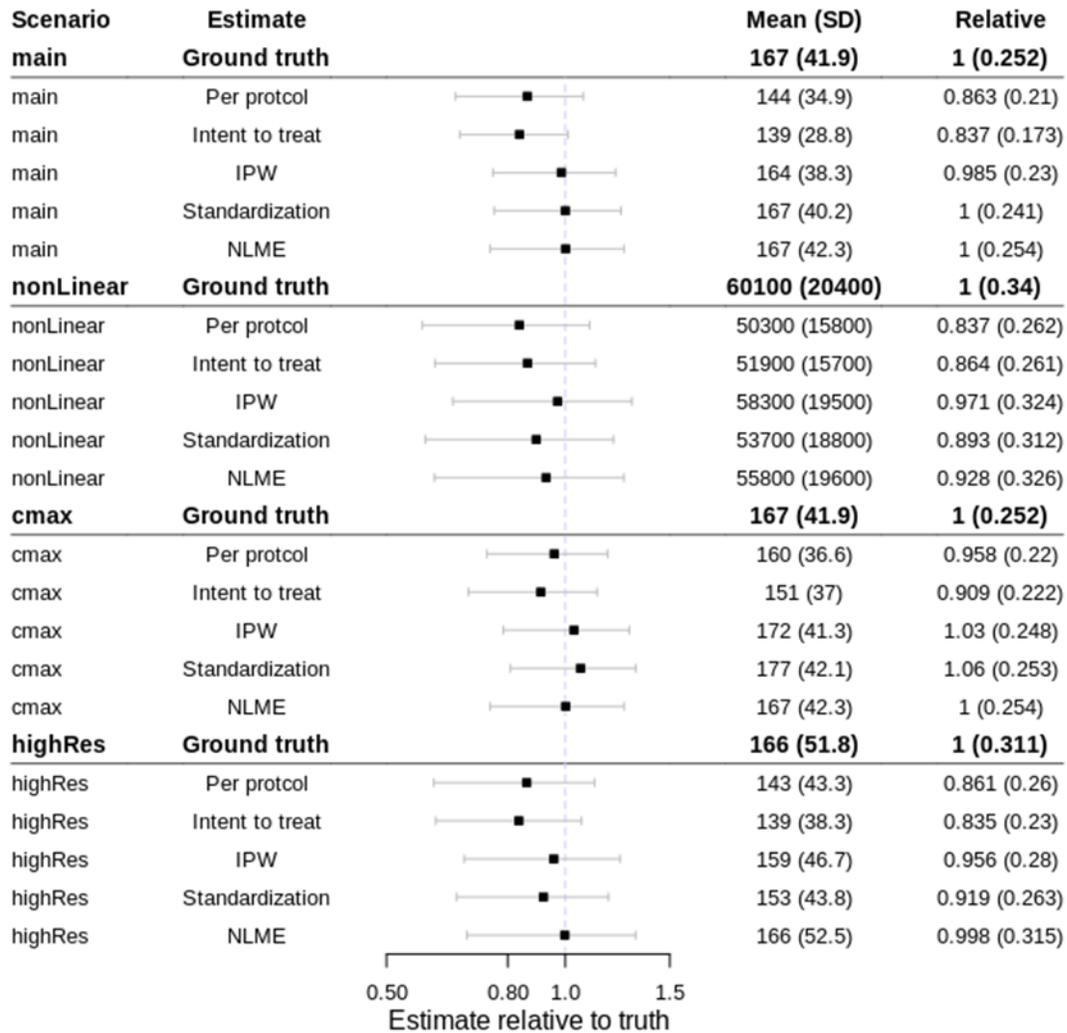

*Mean and standard deviation (SD) of trough exposures at week 8 after the fourth dose for treatment arm 5 (target dose 240, up-titration b). Ground truth and estimates: Standardization, standardization conditioning on previous exposures; NLME, standardization conditioning on individual parameters; IPW, inverse probability weighting. The relative results are the mean and SD of the estimated values divided by the mean of the ground truth. Scenario distinguishes different simulations setups to generate the trial data (ground truth). Main, main scenario used for all other illustrations; nonlinear, assuming non-linear PK with target mediated disposition; cmax, probability of intercurrent events related to maximal concentrations rather than trough concentrations; highRes, residual variability (Table 2) increased to 0.3.*

## Conclusions

NLME M&S can be seen as an implementation of standardization for longitudinal data. Standardization relies on models of the outcome of interest conditional on the confounder. Conventionally, standardization for longitudinal data tends to be implemented sequentially with separate models for subsequent time points that condition on earlier outcomes. In contrast, NLME M&S tends to use a single model that conditions on individual PK parameters (random effects) to describe the evolution of the system over the entire time course. All the g-methods, including non-linear mixed effects modeling, can correct for treatment-confounder feedback, provided that the causal identifiability assumptions hold.

## Acknowledgements

This publication is the result of a thought process that took place over several years. During these years, we discussed aspects and ideas related to this publication with many colleagues. In particular, we want to thank Jean-Louis Steimer and Matt Tudball with whom we started to work on and then abandoned a manuscript with a larger scope of comparing causal inference methods and pharmacometric approaches, in general.